\documentstyle[epsf,axodraw]{article}
\begin{document}

\bigskip
{\Large\bf
	\centerline{QCD Radiative Correction to the Hadronic Annihilation Rate} 
\centerline{of $1^{+-}$ Heavy Quarkonium}
\bigskip
\normalsize

\centerline{Han-Wen Huang$^{1,2}$~~~~Kuang-Ta Chao$^{1,3}$}
\centerline{\sl $^1$ CCAST (World Laboratory), Beijing 100080, P.R.China} 
\centerline{\sl $^2$ Institute of Theoretical Physics, Academia Sinica,
      P.O.Box 2735,}
\centerline{\sl Beijing 100080, P.R.China}
\centerline{\sl $^3$ Department of Physics, Peking University, Beijing 100871,
           P.R.China}
\bigskip

\begin{abstract}
Hadronic annihilation rate of $1^{+-}$ heavy quarkonium is given 
to next-to-leading order in $\alpha_s$ and leading order in $v^2$ 
using a recently developed factorization formalism which is based on NRQCD. 
The result includes both
the annihilation of P-wave color-singlet $Q\bar{Q}$ component, and
the annihilation of S-wave color-octet $Q\bar{Q}$ component of the quarkonium.
The notorious infrared divergences due to soft gluons, i.e., the
Logarithms associated with the binding energy, encountered in previous 
perturbative calculations of $1^{+-}$ quarkonium decays are found to be
explicitly cancelled, and a finite result for the decay width to order
$\alpha_s^3$ is then obtained.
\end{abstract}

\vspace{1cm}
PACS number(s): 12.38.-t, 12.38.Bx, 13.20.Gd 

\vfill\eject\pagestyle{plain}\setcounter{page}{1}

The study of heavy quarkonium plays an important role in the understanding
of quantum chromodynamics (QCD).
In recent years, since the E760 collaboration at Fermilab \cite{e7601,E760}
observed charmonium states via resonant $p\bar{p}$-annihilation and 
measured their masses, widths, and branching fractions with 
unprecedented precision, problems about decay and production of 
heavy quarkonium have aroused much interest of people. 
In particular, with the observation of $1^{+-}$ charmonium state $h_c$
in experiment \cite{e7601}, the study
of its properties such as mass and decay width becomes very interesting both  
theoretically and experimentally. In the 
earlier treatment of heavy quarkonium annihilation, quarkonium is only taken
as a color-singlet bound state of a heavy quark Q and its antiquark $\bar{Q}$.
Calculation of decay rate is based on the assumption that the annihilation 
of Q and $\bar{Q}$
is a short-distance process which, because of asymptotic freedom of QCD,
can be computed in perturbation theory, and all nonperturbative effects could be
factored into a constant: the wavefunction at the origin or its derivative
at the origin. Using this factorization assumption to calculate the hadronic 
decay width of P-wave quarkonium $1^{+-}$ state, the infrared divergence  
appears in the limit of small binding energy \cite{Barbi} and this divergence 
can not be factored into the derivative of the wavefunction.
It is interpreted as a signal that the decay rate is sensitive to the 
nonperturbative effects which can not be contained
in the wavefunction. This implies that the earlier factorization assumption
fails because it is not rigorous within the framework of QCD.

Recently, Bodwin, Braaten and Lepage \cite{BBL} have developed a factorization 
formalism that allows systematic calculations of inclusive decay rates
and production cross section of heavy quarkonium to any order in QCD coupling
constant $\alpha_s$ and to any order in $v^2$, where $v$ is the typical
relative velocity of the heavy quark. The factorization formula is based on 
the use
of the effective field theory NRQCD (Nonrelativistic QCD). Using this rigorous 
factorization formalism, the decay rate can be written as a sum of a set of 
long-distance nonperturbative matrix elements of which each is multiplied by a 
short-distance 
coefficient which can be calculated in perturbative QCD. This approach is
successful in the study of many processes about decay and production of
heavy quarkonium \cite{prod}.
In this paper, we apply NRQCD to the hadronic decay of $1^{+-}$ quarkonium
and compute the QCD radiative corrections to its annihilation rate from
both the color-singlet and color-octet $Q\bar{Q}$ Fock states. We will show
the explicit cancellation of the infrared divergence between the color-singlet
and color-octet $Q\bar{Q}$ components, and give a complete result at next
to leading order in $\alpha_s$ and leading order in $v^2$.

In NRQCD, the effect of annihilation can be taken into account by adding 
4-fermion operators to NRQCD Lagrangian:
\begin{equation}
\delta{\cal L}_{4-fermion}=\sum_n\frac{f_n(\alpha_s)}{m^{d_n-4}}{\cal O}_n,
\end{equation}
where the sum is over all possible local 4-fermion operators ${\cal O}_n$
that annihilate and create a $Q\bar{Q}$ pair, and $d_n$ is the scaling 
dimension of ${\cal O}_n$. The short distance coefficients $f_n(\alpha_s)$ can 
be computed by 
matching perturbative amplitudes for $Q\bar{Q}$ scattering in NRQCD with the 
corresponding amplitudes in full QCD. The annihilation rate of a quarkonium 
state H to light hadrons (LH) can be written as
\begin{equation}
\Gamma(H\rightarrow LH)=2Im<H|\delta{\cal L}_{4-fermion}|H>
\end{equation}
At a given order in $v^2$, the number of matrix elements can be reduced to a 
finite number by using velocity scaling rules for the matrix 
elements\cite{BBL,BBL1}.
These scaling rules consist of that for the operators and 
for the probabilities of the Fock states that give the leading 
contributions to the matrix elements. 

At leading order in $v^2$, the decay
width of $1^{+-}$ quarkonium state can be written as
\begin{equation}
\label{WDA}
\Gamma(1^{+-}\rightarrow LH)=2Imf_1(^1P_1)H_1
+2Imf_8(^1S_0)H_8+O(v^2\Gamma),
\end{equation}
where two nonperturbative parameters $H_1$ and $H_8$ 
can be defined rigorously in terms of matrix elements of a color-singlet
and a color-octet 4-fermion operator in NRQCD
$$
H_1=\frac{<1^{+-}|{\cal O}_1(^1P_1)|1^{+-}>}{m^4},
$$
$$
H_8=\frac{<1^{+-}|{\cal O}_8(^1S_0)|1^{+-}>}{m^2},
$$
where
$$
{\cal O}_1(^1P_1)=\psi^+(-\frac{i}{2}\stackrel{\leftrightarrow}{\bf D})
\chi\cdot\chi^+(-\frac{i}{2}\stackrel{\leftrightarrow}{\bf D})\psi,
$$
$$
{\cal O}_8(^1S_0)=\psi^+T^a\chi\cdot\chi^+T^a \psi,
$$
where ${\bf D}$ is the space part of the covariant derivative $D^{\mu}$ and 
$T^a(a=1,\cdots,N^2_c-1)$ is the $SU(N_c)$ color matrix, and $\psi$ and $\chi^+$
are the fields with two components for quark $Q$ and antiquark $\bar{Q}$ in 
NRQCD.
Here including $H_8$ is due to the fact that for decays of $1^{+-}$ heavy 
quarkonium, an S-wave color-octet $Q\bar{Q}$ component in the wavefunction
will contribute at the same order in $v$ 
as the P-wave color-singlet $Q\bar{Q}$ component,
because the probability for annihilating an S-wave color-octet
$Q\bar{Q}$ state is proportional to $v^0$ while this state has a probability at
order of $v^2$ to be transmitted into a P-wave color-singlet 
$Q\bar{Q}$ state through the emission of soft gluon, whereas the dominate Fock 
state
of $1^{+-}$ quarkonium is $|Q\bar{Q}>$ with $Q\bar{Q}$ pair in a color-singlet
$^1P_1$ state, and the probability for annihilation through P-wave color 
singlet $Q\bar{Q}$ is proportional to $v^2$.

In the following we first calculate the coefficients of non-perturbative 
matrix elements $H_1$ and $H_8$ to order $\alpha_s^3$
by matching the imaginary part of perturbative scattering amplitude of 
$Q\bar{Q}\rightarrow Q\bar{Q}$ in full QCD with that in NRQCD, and then 
derive the formula of $1^{+-}$ quarkonium decay width to an accuracy of next-to-leading order in $\alpha_s$.
Using phenomenological parameters $H_1$ and $H_8$ determined from
other processes, we finally give an approximate numerical estimate of the 
decay width.

We first calculate the imaginary part of $Q\bar{Q}$ forward scattering 
amplitude $Im{\cal M}$ in full QCD. For convenience, we consider
$Q\bar{Q}$ scattering in the center of momentum frame 
with the the momenta of the heavy
quarks and antiquarks small compared to the heavy quark mass.
We take the incoming $Q$ and $\bar{Q}$ to have momenta $\vec{p}$ and 
$-\vec{p}$, while the outgoing Q and $\bar{Q}$ have momenta 
$\vec{p}^{\prime}$ and $-\vec{p}^{\prime}$. By the conservation of energy,
we have $|\vec{p}^{\prime}|=|\vec{p}|\equiv p$. In order to compare
with the result in NRQCD, following \cite{BBL}, in the expression of 
$Im{\cal M}$ to be calculated in full perturbative 
QCD, we write the 4-component Dirac spinors in the Dirac representation in 
terms of 2-component Pauli spinors via the substitutions,  
\begin{eqnarray}
\label{up}
u(\vec p)=\sqrt{\frac{E+m}{2E}}\left(
\begin{array}{c}
\xi\\
\frac{\vec{p}\cdot\vec{\sigma}}{E+m}\xi
\end{array}
\right),
\end{eqnarray}
\begin{eqnarray}
\label{vp}
v(-\vec p)=\sqrt{\frac{E+m}{2E}}\left(
\begin{array}{c}
\frac{-\vec{p}\cdot\vec{\sigma}}{E+m}\eta\\
\eta
\end{array}
\right),
\end{eqnarray}
where $E=\sqrt{m^2+p^2}$, $\xi$ and $\eta$ are 2-component spinors with 
color indices suppressed. The Dirac spinors $u(\vec{p}^{\prime})$
and $v(-\vec{p}^{\prime})$ have similar expressions in terms of Pauli
spinors $\xi^{\prime}$ and $\eta^{\prime}$. The spinors (\ref{up}) and 
(\ref{vp})
represent fermion states with standard nonrelativistic normalization.

\begin{center}\begin{picture}(180,70)(0,0)
\ArrowLine(10,65)(30,65)\ArrowLine(30,65)(30,25)
\ArrowLine(30,25)(10,25)
\ArrowLine(50,65)(70,65)\ArrowLine(50,25)(50,65)
\ArrowLine(70,25)(50,25)
\Photon(30,65)(50,65){1}{5}\Photon(30,25)(50,25){1}{5}
\Text(40,5)[]{(a)}

\ArrowLine(110,65)(130,65)\ArrowLine(130,65)(130,25)
\ArrowLine(130,25)(110,25)
\ArrowLine(150,65)(170,65)\ArrowLine(150,25)(150,65)
\ArrowLine(170,25)(150,25)
\Photon(130,65)(150,25){1}{5}\Photon(130,25)(150,65){1}{5}
\Text(140,5)[]{(b)}
\end{picture}

{\sl\bf Fig.1~~Feynman diagrams contributing to $C^{full~QCD}_{(T^a\otimes T^a) (1\otimes 1)}$ at order $\alpha_s^2$}
\end{center}

It is known from \cite{pwave} that to leading order in $\alpha_s$, only 
the coefficient $Imf_8(^1S_0)$ in (\ref{WDA}) does not vanish and therefore 
only the color-octet matrix element $H_8$ contributes to the decay width.
In the S-wave case, we expand the annihilation amplitude $Im{\cal M}$ 
in terms of velocities $\vec{v}=\vec{p}/E$ and $\vec{v}^{\prime}=\vec{p}^{\prime}/m$ 
only to leading order, and reduce $Im{\cal M}$ to four terms
\begin{eqnarray}\nonumber
&&Im{\cal M}\\\nonumber&=&
C_{(1\otimes 1)(1\otimes 1)}\xi^{\prime +}\eta^{\prime}\eta^+\xi
+C_{(T^a\otimes T^a)(1\otimes 1)} 
\xi^{\prime +}T^a\eta^{\prime}\eta^+T^a\xi\\
&+&C_{(1\otimes 1)(\sigma^i\otimes \sigma^i)} 
\xi^{\prime +}\vec{\sigma}\eta^{\prime}\cdot\eta^+\vec{\sigma}\xi
+C_{(T^a\otimes T^a)(\sigma^i\otimes \sigma^i)} 
\xi^{\prime +}\vec{\sigma}T^a\eta^{\prime}\cdot\eta^+\vec{\sigma}T^a\xi.
\end{eqnarray}
In order to determine 
$(Imf_8(^1S_0))_0$, we only consider the coefficient 
$C_{(T^a\otimes T^a)(1\otimes 1)}$ 
of the term $\xi^{\prime +}
T^a\eta^{\prime}\eta^+T^a\xi$. At order $\alpha_s^2$, 
only two diagrams shown in Fig.1(a) and Fig.1(b) contribute to this coefficient. 
After decomposing the spinors and expanding them
to leading order in $\vec{v}$ and $\vec{v}^{\prime}$, 
we obtain the coefficient of the term $\xi^{\prime +}
T^a\eta^{\prime}\eta^+T^a\xi$ in $Im{\cal M}$
\begin{equation}
C^{full~QCD}_{(T^a\otimes T^a) (1\otimes 1)}=\frac{(N_c^2-4)g^4}{16N_cm^2}
(d-2)(d-3)\Phi(2), 
\end{equation}
where the two massless particle phase space $\Phi(2)$ in $d=4-2\epsilon$ 
dimension is integrated to give
$$
\Phi(2)=\frac{1}{8\pi}(\frac{4\pi}{4m^2})^{\epsilon}\frac{\Gamma(1-\epsilon)}
{\Gamma(2-2\epsilon)}.
$$
A simple calculation leads to the final expression 
\begin{equation}
\label{c0}
C^{full~QCD}_{(T^a\otimes T^a) (1\otimes 1)}=\frac{\pi(N_c^2-4)\alpha_s^2}
{4N_cm^2}
(\frac{4\pi\mu^4}{4m^2})^{\epsilon}\frac{\Gamma(1-\epsilon)}{\Gamma
(2-2\epsilon)}(1-\epsilon)(1-2\epsilon),
\end{equation}
where the dimensionless coupling constant is defined as
$$
\alpha_s=(\frac{g^2}{4\pi})\mu^{-2\epsilon}.
$$

While in NRQCD the $Q\bar{Q}$ forward scattering amplitude can be reproduced 
by 4-fermion
operators in the effective lagrangian, and the corresponding term 
$\xi^{\prime +}T^a\eta^{\prime}\eta^+T^a\xi$ in $Im{\cal M}$ comes from
operator $O_8(^1S_0)$, of which the coefficient is $\frac{Imf_8(^1S_0)}{m^2}$. 
By comparing it with (\ref{c0}),
we get
\begin{equation}\label{f80}
(Imf_8(^1S_0))_0=\frac{\pi(N_c^2-4)\alpha_s^2}{4N_c}
(\frac{4\pi\mu^4}{4m^2})^{\epsilon}\frac{\Gamma(1-\epsilon)}{\Gamma
(2-2\epsilon)}(1-\epsilon)(1-2\epsilon).
\end{equation}
We keep $\epsilon$ in the above expression for convenience of later 
calculations.
In the limit $\epsilon\rightarrow 0$, we get 
\begin{equation}
(Imf_8(^1S_0))_0=\frac{\pi(N_c^2-4)}{4N_c}\alpha_s^2,
\end{equation}
which has been given in \cite{pwave}.
Here the subscript $``0"$ in the coefficient means only the result at
leading order in $\alpha_s$ is taken, and the width can be written as
\begin{equation}
\Gamma(1^{+-}\rightarrow LH)=\frac{\pi(N_c^2-4)}{2N_c}\alpha_s^2H_8
+O(\alpha_s\Gamma).
\end{equation}

In order to obtain the next-to-leading order result, we must take account of 
the effects coming from both color-octet component and color-singlet 
component of the quarkonium.
In the following, we calculate the coefficients $Imf_1(^1P_1)$ to leading order
in $\alpha_s$ and $Imf_8(^1S_0)$ to next-to-leading order in $\alpha_s$,
and then give the complete formular for the hadronic decay width of $^1P_1$ 
to order of $\alpha_s^3$ at leading order of $v^2$. 

\begin{center}\begin{picture}(230,60)(0,0)
\Line(0,50)(15,50)\Line(0,20)(15,20)\Line(15,20)(15,50)
\Photon(15,50)(35,50){1}{6}\Photon(15,35)(35,35){1}{6}
\Photon(15,20)(35,20){1}{6}
\Line(35,50)(50,50)\Line(35,20)(50,20)\Line(35,20)(35,50)
\Text(25,10)[]{(a)}

\Line(60,50)(75,50)\Line(60,20)(75,20)\Line(75,20)(75,50)
\Photon(75,50)(95,50){1}{6}\Photon(75,35)(95,20){1}{8}
\Photon(75,20)(95,35){1}{8}
\Line(95,50)(110,50)\Line(95,20)(110,20)\Line(95,20)(95,50)
\Text(85,10)[]{(b)}

\Line(120,50)(135,50)\Line(120,20)(135,20)\Line(135,20)(135,50)
\Photon(135,50)(155,35){1}{8}\Photon(135,35)(155,20){1}{8}
\Photon(135,20)(155,50){1}{8}
\Line(155,50)(170,50)\Line(155,20)(170,20)\Line(155,20)(155,50)
\Text(145,10)[]{(c)}

\Line(180,50)(195,50)\Line(180,20)(195,20)\Line(195,20)(195,50)
\Photon(195,50)(215,20){1}{8}\Photon(195,35)(215,35){1}{8}
\Photon(195,20)(215,50){1}{8}
\Line(215,50)(230,50)\Line(215,20)(230,20)\Line(215,20)(215,50)
\Text(205,10)[]{(d)}
\end{picture}

{\sl\bf Fig.2~~Feynman diagrams contributing to $C^{full~QCD}_{\vec{v}^{\prime}\cdot\vec{v}(1\otimes 1)(1\otimes 1)}$ at leading order in $\alpha_s$}
\end{center}

Via the same procedure as above, we consider the imaginary part of $Q\bar{Q}$ 
scattering amplitude, and calculate the coefficient $Imf_1(^1P_1)$ by
matching a perturbative calculation in full QCD with the corresponding
perturbative calculation in NRQCD.  
The P-wave case requires an expansion of the annihilation
amplitude $Im{\cal M}$ up to the first power of relative momenta $\vec{p}$
and $\vec{p}^{\prime}$. At this order $Im{\cal M}$ can be written as
\begin{eqnarray}\nonumber
&&Im{\cal M}\\\nonumber&=&
C_{\vec{v}^{\prime}\cdot\vec{v}(1\otimes 1)(1\otimes 1)} 
\vec{v}^{\prime}\cdot\vec{v}\xi^{\prime +}
\eta^{\prime}\eta^+\xi+
C_{\vec{v}^{\prime}\cdot\vec{v}(1\otimes 1)(\sigma^i\otimes\sigma^i)} 
\vec{v}^{\prime}\cdot\vec{v}\xi^{\prime +}\vec{\sigma}
\eta^{\prime}\cdot\eta^+\vec{\sigma}\xi\\\nonumber
&+&
C_{v^{\prime i}v^j(1\otimes 1)(\sigma^i\otimes\sigma^j)} 
\xi^{\prime +}\vec{v}^{\prime}\cdot\vec{\sigma}
\eta^{\prime}\eta^+\vec{v}\cdot\vec{\sigma}\xi+
C_{v^{\prime j}v^i(1\otimes 1)(\sigma^i\otimes\sigma^j)} 
\xi^{\prime +}\vec{v}\cdot\vec{\sigma}
\eta^{\prime}\eta^+\vec{v}^{\prime}\cdot\vec{\sigma}\xi\\\nonumber
&+&C_{\vec{v}^{\prime}\cdot\vec{v}(T^a\otimes T^a)(1\otimes 1)} 
\vec{v}^{\prime}\cdot\vec{v}\xi^{\prime +}T^a
\eta^{\prime}\eta^+T^a\xi+
C_{\vec{v}^{\prime}\cdot\vec{v}(T^a\otimes T^a)(\sigma^i\otimes\sigma^i)} 
\vec{v}^{\prime}\cdot\vec{v}\xi^{\prime +}\vec{\sigma}T^a
\eta^{\prime}\cdot\eta^+\vec{\sigma}T^a\xi\\
&+&C_{v^{\prime i}v^j(T^a\otimes T^a)(\sigma^i\otimes\sigma^j)} 
\xi^{\prime +}\vec{v}^{\prime}\cdot\vec{\sigma}T^a
\eta^{\prime}\eta^+\vec{v}\cdot\vec{\sigma}T^a\xi+
C_{v^{\prime j}v^i(T^a\otimes T^a)(\sigma^i\otimes\sigma^j)} 
\xi^{\prime +}\vec{v}\cdot\vec{\sigma}T^a
\eta^{\prime}\eta^+\vec{v}^{\prime}\cdot\vec{\sigma}T^a\xi.
\end{eqnarray}
The determination of $Imf_1(^1P_1)$
only requires calculating the coefficient 
$C_{\vec{v}^{\prime}\cdot\vec{v}(1\otimes 1)(1\otimes 1)}$ 
of the term~~ 
$\vec{v}^{\prime}\cdot\vec{v}\xi^{\prime +}
\eta^{\prime}\eta^+\xi$. In full QCD to leading order
in $\alpha_s$, only the diagrams in Fig.2 contribute to this coefficient. After
making a nonrelativistic expansion for $Im{\cal M}$ to first order
in $v$ and $v^{\prime}$, we get
\begin{eqnarray}\nonumber
&&C^{full~QCD}_{\vec{v}^{\prime}\cdot\vec{v}(1\otimes 1)(1\otimes 1)}\\\nonumber
&=&\int\frac{(N_c^2-4)C_Fg^6}{8N_c^2m^4}\frac{d-3}{48(d-1)}\{[
(-320+96d)\frac{1}{x_1^3x_2^3}+(768-240d+4d^2)(\frac{1}{x^3_1x^2_2}
+\frac{1}{x_1^2x_2^3})\\\nonumber
&+&(-512+176d-4d^2)(\frac{1}{x^3_1x_2}+\frac{1}{x_1x^3_2})
+(64-32d)(\frac{1}{x_1^3}+\frac{1}{x_2^3})\\\nonumber
&+&(-1344+404d-2d^2)\frac{1}{x^2_1x^2_2}+(624-208d+2d^2)
(\frac{1}{x^2_1x_2}+\frac{1}{x_1x^2_2})\\\nonumber
&+&(-64+32d)(\frac{1}{x_1^2}+\frac{1}{x_2^2})+(-168+68d-3d^2)\frac{1}{x_1x_2}]\\
&+&(two~~other~~permutations)\}d\Phi(3).
\end{eqnarray}
Here $x_i=k_i/m~(i=1,2,3)$ and $k_i$ denote the energies of the final-state
gluons. The massless three-body phase space can be written as
$$
d\Phi(3)=\frac{4m^2}{2(4\pi)^3}(\frac{4\pi}{4m^2})^{2\epsilon}
\frac{1}{\Gamma(2-2\epsilon)}[(1-x_1)(1-x_2)(1-x_3)]^{-\epsilon}
dx_1dx_2.
$$
After performing the integration for two invariant $x_1$ and $x_2$, 
we obtain  
\begin{eqnarray}\label{pw0}\nonumber
C^{full~QCD}_{\vec{v}^{\prime}\cdot\vec{v}(1\otimes 1)(1\otimes 1)}&=&
\frac{(N_c^2-4)C_F\alpha_s^3}{3N_c^2m^2}(\frac{4\pi}{4m^2})
^{2\epsilon}(\mu^{2\epsilon})^3\frac{1-2\epsilon}{\Gamma(2-
2\epsilon)}(-\frac{1}{2\epsilon_{IR}}+\frac{7\pi^2-94}{48})\\
&=&\frac{(Imf_8(^1S_0))_0}{m^2}\frac{4C_F\alpha_s}{3N_c\pi}
[-\frac{1}{2}(\frac{1}{\epsilon_{IR}}-\gamma_E+ln\frac{4\pi\mu_{IR}^2}
{4m^2})+\frac{7\pi^2-118}{48}].
\end{eqnarray}
where $\gamma_E=0.577$ is the Euler constant. 
Comparing (\ref{pw0}) with the result obtained in ref.\cite{Barbi} which is 
regularized by the binding energy of $Q\bar{Q}$
pair, we find that if making the substitution $ln\frac{m}{\varepsilon}
\rightarrow -\frac{1}{2\epsilon_{IR}}$, the two results have the same 
divergent terms,
but their finite terms are different due to different regularization schemes.
Here $\varepsilon$ is the binding energy of 
$Q\bar{Q}$ pair, which is defined as
$$
\frac{\varepsilon}{m}=\frac{4m^2-M^2}{4m^2},
$$
where $M$ is the mass of $Q\bar{Q}$ bound state. Here we control the infrared 
divergence using on-shell dimensional
regularization, because off-shell binding energy regularization
scheme will break manifest gauge invariance and conventional treatment of
NRQCD is exact only for on-shell amplitudes. However, after taking account of
the contribution from color-octet $Q\bar{Q}$ component 
we will find that the coefficient $Imf_1(^1P_1)$ is infrared finite and the 
final result is independent of infrared regularization scheme.

\begin{center}\begin{picture}(300,60)(0,0)
\ArrowLine(100,50)(150,30)\ArrowLine(200,10)(150,30)
\ArrowLine(150,30)(100,10)\ArrowLine(150,30)(200,50)
\Vertex(150,30){2}
\end{picture}

{\sl\bf Fig.3~~Feynman diagram contributing to $C^{NRQCD}_{\vec{v}^{\prime}\cdot\vec{v}(1\otimes 1)(1\otimes 1)}$
 through the operator ${\cal O}_1(^1P_1)$}
\end{center}

\begin{center}\begin{picture}(300,120)(0,0)
\ArrowLine(10,110)(75,90)\ArrowLine(140,70)(75,90)
\ArrowLine(75,90)(140,110)\ArrowLine(75,90)(10,70)
\Photon(42.5,100)(107.5,100){1}{10}
\Vertex(75,90){2}
\Text(75,65)[]{(a)}

\ArrowLine(160,110)(225,90)\ArrowLine(290,70)(225,90)
\ArrowLine(225,90)(290,110)\ArrowLine(225,90)(160,70)
\Photon(192.5,80)(257.5,80){1}{10}
\Vertex(225,90){2}
\Text(225,65)[]{(b)}

\ArrowLine(10,50)(75,30)\ArrowLine(140,10)(75,30)
\ArrowLine(75,30)(140,50)\ArrowLine(75,30)(10,10)
\PhotonArc(75,30)(20,344,164){1}{20}
\Vertex(75,30){2}
\Text(75,5)[]{(c)}

\ArrowLine(160,50)(225,30)\ArrowLine(290,10)(225,30)
\ArrowLine(225,30)(290,50)\ArrowLine(225,30)(160,10)
\PhotonArc(225,30)(20,16,196){1}{20}
\Vertex(225,30){2}
\Text(225,5)[]{(d)}
\end{picture} 

{\sl\bf Fig.4~~Feynman diagrams contributing to $C^{NRQCD}_{\vec{v}^{\prime}\cdot\vec{v}(1\otimes 1)(1\otimes 1)}$
through the operator ${\cal O}_8(^1S_0)$}
\end{center}

In NRQCD, the $Q\bar{Q}$ forward scattering amplitude can be reproduced by 
operators in $\delta{\cal L}_{4-fermion}$. When working at order $\alpha_s^3$,
there are two 4-fermion operators which contribute to the coefficient 
$C_{\vec{v}^{\prime}\cdot\vec{v}(1\otimes 1)(1\otimes 1)}$ of the term 
$\vec{v}^{\prime}\cdot\vec{v}\xi^{\prime +}\eta^{\prime}\eta^+\xi$ in 
$Im{\cal M}$, which are
$$
\delta{\cal L}_{4-fermion}=\frac{f_1(^1P_1)}{m^4}{\cal O}_1(^1P_1)
+\frac{f_8(^1S_0)}{m^2}{\cal O}_8(^1S_0).
$$
The color singlet operator ${\cal O}_1(^1P_1)$ contributes through the tree  
diagram in Fig.3 which contains a 4-fermion vertex corresponding to ${\cal O}_1(^1P_1)$
, and the result is
\begin{equation}
Im{\cal M}_{Fig.3}=\frac{Imf_1(^1P_1)}{m^2}\vec{v}^{\prime}\cdot\vec{v}
\xi^{\prime +}\eta^{\prime}\eta^+\xi.
\end{equation}
Since $Imf_8(^1S_0)$ is already known to be of order $\alpha_s^2$, it is
necessary to compute the contribution of
 the operator ${\cal O}_8(^1S_0)$ to an accuracy of $\alpha_s$. It is obvious
that this contribution only comes 
from one-loop diagrams in Fig.4(a)--(d) which 
contain a 4-fermion vertex corresponding to ${\cal O}_8(^1S_0)$,
and these one-loop figures cause the transition from a color 
octet $Q\bar{Q}$ into a color singlet $Q\bar{Q}$. 
The overall contribution of diagrams in Fig.4 is
\begin{eqnarray}\label{mid}\nonumber
Im{\cal M}_{Fig.4}&=&\frac{Imf_8(^1S_0)}{m^2}\frac{4C_F\alpha_s}{3N_c\pi}
[-\frac{1}{2}(\frac{1}{\epsilon_{IR}}-\gamma_E+ln\frac{4\pi\mu_{IR}^2}
{4m^2})\\
&+&\frac{1}{2}(\frac{1}{\epsilon_{UV}}-\gamma_E+ln\frac{4\pi\mu_{UV}^2}
{4m^2})]\vec{v}^{\prime}\cdot\vec{v}
\xi^{\prime +}\eta^{\prime}\eta^+\xi,
\end{eqnarray}
where $\frac{1}{\epsilon_{IR}}$ is the IR (infrared) divergence and $\mu_{IR}$ 
is the corresponding scale, while $\frac{1}{\epsilon_{UV}}$ is the UV 
(ultraviolet) divergence and $\mu_{UV}$ is 
the corresponding scale. After the renormalization of 
operator ${\cal O}_8(^1S_0)$ in the $\overline{MS}$ scheme the result is free 
from UV divergence, but the IR
divergence still remains and it represents the nonperturbative nature
of the annihilation amplitude. To order $\alpha_s^3$, $Imf_8(^1S_0)$
on the right hand side of (\ref{mid}) must be taken as $(Imf_8(^1S_0))_0$, 
and then we obtain
\begin{equation}
\label{pw}
C^{NRQCD}_{\vec{v}^{\prime}\cdot\vec{v}(1\otimes 1)(1\otimes 1)}=
\frac{Imf_1(^1P_1)}{m^2}+\frac{(Imf_8(^1S_0))_0}{m^2}\frac{4C_F\alpha_s}
{3N_c\pi}
[-\frac{1}{2}(\frac{1}{\epsilon_{IR}}-\gamma_E+ln\frac{4\pi\mu_{IR}^2}
{4m^2})+ln\frac{\mu_{UV}}{2m})].
\end{equation}
From (\ref{pw0}) and (\ref{pw}), we find that the coefficients
of IR divergence are the same. It is clear that the IR divergence appearing
in (\ref{pw0}) is proportional to the probability of transition between
a color-singlet $Q\bar{Q}$ pair and a color-octet $Q\bar{Q}$ pair by the
emission of soft gluon. This is the nonperturbative effect and must be factored
into the long-distance matrix elements which have been defined explicitly in 
NRQCD. Comparing (\ref{pw0}) with (\ref{pw}) and using (\ref{f80}),
the finite coefficient $Imf_1(^1P_1)$ is found to be
\begin{equation}
\label{IMP}
Imf_1(^1P_1)=\frac{(N_c^2-4)C_F\alpha_s^3}{3N_c^2}(\frac{7\pi^2-118}{48}
-ln\frac{\mu}{2m}).
\end{equation}
Obviously the previously encountered IR divergence has been canceled and 
factored into the nonperturbative matrix element. The operator 
${\cal O}_8(^1S_0)$ satisfies the evolution equation
\begin{equation}
\mu\frac{\partial{\cal O}_8(^1S_0)}{\partial\mu}=\alpha_s(\mu)\frac{4C_F}
{3\pi N_cm^2}{\cal O}_1(^1P_1),
\end{equation}
which has been derived in \cite{BBL}. We have neglected the subscript $``UV"$ 
in $\mu$ and we will keep this notation in our work.

\begin{center}\begin{picture}(250,400)(0,0)
\Line(10,380)(30,380)\Line(10,340)(30,340)
\Line(30,340)(30,380)\Photon(20,340)(20,380){1}{8}
\Photon(30,380)(60,380){1}{6}\Photon(30,340)(60,340){1}{6}
\Line(60,340)(80,340)\Line(60,380)(80,380)
\Line(60,340)(60,380)
\Text(45,325)[]{(a)}

\Line(90,380)(110,380)\Line(90,340)(110,340)
\Line(110,340)(110,380)\PhotonArc(110,360)(10,90,270){1}{6}
\Photon(110,380)(140,380){1}{6}\Photon(110,340)(140,340){1}{6}
\Line(140,380)(160,380)\Line(140,340)(160,340)
\Line(140,380)(140,340)
\Text(125,325)[]{(b)}

\Line(170,380)(190,380)\Line(170,340)(190,340)
\Line(190,340)(190,380)
\PhotonArc(180,380)(8,180,360){1}{6}
\Photon(190,340)(220,340){1}{6}\Photon(190,380)(220,380){1}{6}
\Line(220,340)(240,340)\Line(220,380)(240,380)
\Line(220,340)(220,380)
\Text(205,325)[]{(c)}

\Line(10,310)(30,310)\Line(10,270)(30,270)
\Line(30,310)(30,270)\PhotonArc(30,310)(15,180,270){1}{6}
\Photon(30,310)(60,310){1}{6}\Photon(30,270)(60,270){1}{6}
\Line(60,310)(80,310)\Line(60,270)(80,270)
\Line(60,310)(60,270)
\Text(45,255)[]{(d)}
 
\Line(90,310)(110,310)\Line(90,270)(110,270)\Line(110,270)(110,310)
\Photon(110,310)(140,310){1}{6}\Photon(110,270)(140,270){1}{6}
\Photon(110,290)(125,310){1}{4}\Vertex(125,310){1}
\Line(140,310)(160,310)\Line(140,270)(160,270)\Line(140,270)(140,310)
\Text(125,255)[]{(e)}

\Line(170,310)(190,310)\Line(170,270)(190,270)\Line(190,270)(190,310)
\Photon(190,310)(200,310){1}{2}\Vertex(205,310){5}
\Photon(210,310)(220,310){1}{2}\Photon(190,270)(220,270){1}{6}
\Line(220,310)(240,310)\Line(220,270)(240,270)\Line(220,270)(220,310)
\Text(205,255)[]{(f)}

\Line(10,240)(30,240)\Line(10,200)(30,200)\Line(30,240)(30,200)
\Photon(30,240)(40,240){1}{2}\CArc(45,240)(5,0,360)
\Photon(50,240)(60,240){1}{2}\Photon(30,200)(60,200){1}{6}
\Line(60,240)(80,240)\Line(60,200)(80,200)\Line(60,200)(60,240)
\Text(45,185)[]{(g)}

\Line(90,240)(110,240)\Line(90,200)(110,200)\Line(110,200)(110,240)
\Photon(110,240)(140,200){1}{8}\Photon(110,200)(140,240){1}{8}
\Vertex(125,220){1}
\Line(140,240)(160,240)\Line(140,200)(160,200)\Line(140,200)(140,240)
\Text(125,185)[]{(h)}

\Line(170,240)(190,240)\Line(170,200)(190,200)\Line(190,200)(190,240)
\Photon(190,240)(220,240){1}{6}\Photon(190,200)(205,240){1}{8}
\Photon(190,220)(220,200){1}{8}
\Line(220,240)(240,240)\Line(220,200)(240,200)\Line(220,200)(220,240)
\Text(205,185)[]{(i)}
\Line(10,170)(30,170)\Line(10,130)(30,130)\Line(30,130)(30,170)
\Photon(30,170)(60,170){1}{6}\Photon(30,130)(60,130){1}{6}
\Photon(45,170)(45,130){1}{8}
\Line(60,170)(80,170)\Line(60,130)(80,130)\Line(60,130)(60,170)
\Text(45,115)[]{(j)}

\Line(90,170)(110,170)\Line(90,130)(110,130)\Line(110,130)(110,170)
\Photon(110,170)(140,170){1}{6}\Photon(110,150)(140,150){1}{6}
\Photon(110,130)(140,130){1}{6}
\Line(140,130)(160,130)\Line(140,170)(160,170)\Line(140,130)(140,170)
\Text(125,115)[]{(k)}

\Line(170,170)(190,170)\Line(170,130)(190,130)\Line(190,130)(190,170)
\Photon(190,170)(220,130){1}{8}\Photon(190,150)(220,150){1}{6}
\Photon(190,130)(220,170){1}{8}
\Line(220,170)(240,170)\Line(220,130)(240,130)\Line(220,130)(220,170)
\Text(205,115)[]{(l)}

\Line(10,100)(30,100)\Line(10,60)(30,60)\Line(30,60)(30,100)
\Photon(30,100)(60,100){1}{6}\Photon(30,80)(60,60){1}{8}
\Photon(30,60)(60,80){1}{8}
\Line(60,100)(80,100)\Line(60,60)(80,60)\Line(60,60)(60,100)
\Text(45,45)[]{(m)}

\Line(90,100)(110,100)\Line(90,60)(110,60)\Line(110,60)(110,100)
\Photon(110,100)(140,80){1}{8}\Photon(110,80)(140,60){1}{8}
\Photon(110,60)(140,100){1}{8}
\Line(140,100)(160,100)\Line(140,60)(160,60)\Line(140,60)(140,100)
\Text(125,45)[]{(n)}

\Photon(10,20)(20,20){1}{2}\Vertex(25,20){5}
\Photon(30,20)(40,20){1}{2}\Text(50,20)[]{=}
\Photon(60,20)(70,20){1}{2}
\PhotonArc(75,20)(5,0,360){1}{8}
\Photon(80,20)(90,20){1}{2}
\Text(95,20)[]{+}
\Photon(100,20)(115,20){1}{3}
\PhotonArc(115,25)(5,0,360){1}{8}
\Photon(115,20)(130,20){1}{3}
\Text(135,20)[]{+}
\Photon(140,20)(150,20){1}{2}
\DashCArc(155,20)(5,0,360){0.4}
\Photon(160,20)(170,20){1}{2}
\end{picture}

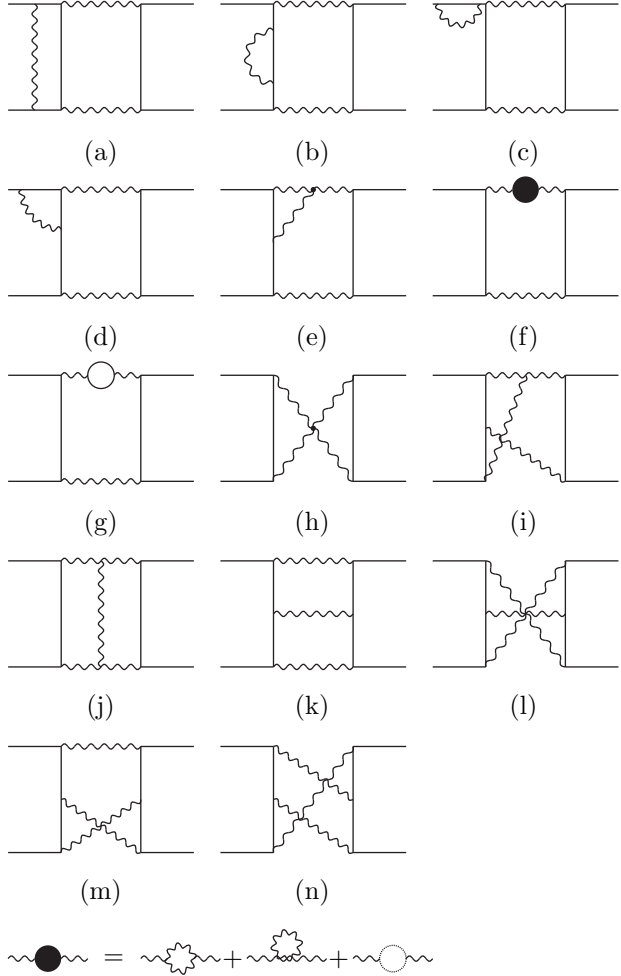
{\sl\bf Fig.5~~Representive diagrams 
 contributing to the first-order radiative correction to $C^{full~QCD}_{(T^a\otimes T^a) (1\otimes 1)}$ }
\end{center}

We have derived the coefficient $Imf_8(^1S_0)$ to leading order in
$\alpha_s$. In order to get the result to next-to-leading order, we
must consider the imaginary part of scattering amplitude of $Q\bar{Q}$ pair 
to order in $\alpha_s^3$ in full QCD.
The diagrams which contribute to the coefficient of the term
$\xi^{\prime +}T^a\eta^{\prime}\eta^+T^a\xi$ in $Im{\cal M}$ 
to next-to-leading order in $\alpha_s$ are shown in Fig.5.
We only give the representative diagrams and neglect the diagrams which
give the same result as some of those in Fig.5. The contribution from each diagram in
terms of the unrenormalized coupling constant has in general the following form:
\begin{equation}
\frac{(Imf_8(^1S_0))_0}{m^2}\frac{\alpha_s}{\pi}f(\epsilon)A("diagram")
\end{equation}
with
$$
f(\epsilon)=(\frac{4\pi\mu^2}{4m^2})^{\epsilon}\Gamma(1+\epsilon).
$$
The imaginary part of these diagrams receives contributions from two-gluon
cut, three-gluon cut and a ``light" quark-antiquark pair plus one-gluon cut.
The contribution of each individual diagram is
calculated in Feynman gauge. Hence we have to add a ghost contribution 
both to two-gluon cut and to three-gluon cut in the diagram of Fig.5f.
Our results for the contributions from individual diagrams are listed in 
table 1. 

\begin{table}
\begin{center}

\caption{$C^{full~QCD}_{(T^a\otimes T^a) (1\otimes 1)}$
from individual diagrams shown in Fig.5}

\begin{tabular}{|c|c|c|}\hline\hline
& Contribution from & Contribution from \\
Diagram  &two-particle cut& three-particle cut \\\hline
(a)&$(C_F-\frac{C_A}{2})(\frac{\pi^2}{2v}+\frac{1}{\epsilon_{IR}}+
2ln2-2)$&0\\\hline
(b)&$C_F(-\frac{1}{2\epsilon_{UV}}+3ln2-1)$&0\\\hline
(c)&$C_F(-\frac{1}{2\epsilon_{UV}}-\frac{1}{\epsilon_{IR}}-3ln2-2)$&0\\\hline
(d)&$(C_F-\frac{C_A}{2})(\frac{1}{\epsilon_{UV}}-2ln2+\frac{\pi^2}{4})$&0
\\\hline
(e)&$C_A[\frac{3}{2\epsilon_{UV}}-\frac{1}{2}(\frac{1}{\epsilon^2}
+\frac{1}{\epsilon})_{IR}+2-ln2+\frac{\pi^2}{12}]$
&$C_A[\frac{1}{2}(\frac{1}{\epsilon^2}+\frac{1}{\epsilon})_{IR}+\frac{1}{2}
-\frac{\pi^2}{6}]$\\\hline
(f)&$C_A(\frac{5}{6\epsilon_{UV}}-\frac{5}{6\epsilon_{IR}})$
&$C_A(\frac{5}{6\epsilon_{IR}}+\frac{23}{9})$\\\hline
(g)&$-\frac{n_f}{3\epsilon_{UV}}+\frac{2}{3}\sum_iln\frac{m_i}{2m}$&
$-\frac{2}{3}\sum_iln\frac{m_i}{2m}-\frac{8n_f}{9}$\\\hline
(h)&0&0\\\hline
(i)&0&0\\\hline
(j)&$\frac{C_A}{2}[-(\frac{1}{\epsilon^2}+\frac{1}{\epsilon})_{IR}
-2+2ln2+\frac{2\pi^2}{3}]$&$\frac{C_A}{2}[(\frac{1}{\epsilon^2}
+\frac{1}{\epsilon})_{IR}+9-\frac{4\pi^2}{3}]$\\\hline
$(k)+(l)$&0&
$\frac{C_A}{2}(\frac{1}{\epsilon_{IR}}+8-\frac{\pi^2}{3})$\\\hline
$(m)+(n)$&0&0\\\hline\hline
\end{tabular}

\centerline{The values are normalized to $\frac{(Imf_8(^1S_0))_0}{m^2}
f(\epsilon)\frac{\alpha_s}{\pi}$.}

\centerline{Here $\epsilon=(4-d)/2$,
~~$f(\epsilon)=(\frac{4\pi\mu^2}{4m^2})^{\epsilon}\Gamma(1+\epsilon)$,~~$
C_F=\frac{N_c^2-1}{2N_c},~~C_A=N_c$,}

\centerline{~~$n_f$ stands for the number of 
light flavors}

\end{center}
\end{table}

Divergences show up in the intermediate steps of the calculation, the
dimensional regularization procedure is used by going to $d$ dimensions and 
introducing a scale $\mu$ through the standard replacement of the bare coupling
constant $g\rightarrow g\mu^{(d-4)/2}$. 
Manifest gauge invariance and massless particle kinematics greatly simplify the
calculations. The origin of the $\epsilon=0$ poles
is specified in the table by the subscripts UV and IR. In the table we
give the regularized und nrenormalized results for these diagrams, which
show a $1/(d-4)$ divergence and a finite part.

The overall result for the unrenormalized first-order radiative correction to 
the coefficient $C_{(T^a\otimes T^a) (1\otimes 1)}$ in full QCD can be 
obtained by summing up all different individual contributions, and reads
\begin{eqnarray}\nonumber
&&C^{full~QCD}_{(T^a\otimes T^a) (1\otimes 1)}\\
&=&\frac{(Imf_8(^1S_0))_0}{m^2}
\{1+\frac{\alpha_s}{\pi}f(\epsilon)[2b_0\frac{1}{\epsilon}
+(C_F-\frac{C_A}{2})\frac{\pi^2}{2v}+A]\},
\end{eqnarray}
where
$$
f(\epsilon)=(\frac{4\pi\mu^2}{4m^2})^{\epsilon}\Gamma(1+\epsilon),
$$
$$
b_0=\frac{1}{12}(11C_A-2n_f),
$$
$$
A=C_F(\frac{\pi^2}{4}-5)
+C_A(\frac{122}{9}\footnote{I would like to thank Dr. Petrelli $et~al.$
to point out the problem of this term for our old version in their
recent paper hep-ph/9707223. Now in this new version it is important to
note that the two independent calculations are found to give a identical
result.}-\frac{17\pi^2}{24})-\frac{8}{9}n_f,
$$
with $C_F=\frac{N_c^2-1}{2N_c}$,~~$C_A=N_c$. The term with $\frac{1}{v}$ 
indicates the 
Coulomb singularity which arises from the Coulomb exchange of the gluon between
quark and antiquark in Fig.5(a). It is sensitive to the long-distance
nonperturbative effect. 

We find that the cancellation of the infrared divergences occurs in the 
overall result
in spite of the fact that the individual cuts do not. This
is the same as the corresponding color singlet coefficient $C^{full~QCD}_
{(1\otimes 1)(1\otimes 1)}$. As a matter of fact it may be 
interesting to know that the infrared divergences of Fig.5a,c,k,l cancel each 
other, which is different from the color singlet coefficient, where the 
cancellation occurs between Fig.5a and Fig.5c, as well as between Fig.5k, 
Fig.5l, Fig.5m and Fig.5n respectively \cite{Barbi1,Hag}. 

Now we renormalize the coupling constant in the $\overline{MS}$ scheme with
$$
\frac{\alpha_s}{\pi}=\frac{\alpha_s^{\overline{MS}}}{\pi}
(1-\frac{\alpha_s^{\overline{MS}}}{\pi}b_0(\frac{1}{\epsilon}
+ln4\pi-\gamma_E)),
$$
and find
\begin{eqnarray}\nonumber
\label{PES}
&&C^{full~QCD}_{(T^a\otimes T^a) (1\otimes 1)}\\&=
&\frac{\pi(N_c^2-4)}{4N_cm^2}\alpha_s^2\{1
+\frac{\alpha_s}{\pi}[(C_F-\frac{C_A}{2})\frac{\pi^2}{2v}
+4b_0ln\frac{\mu}{2m}+A]\},
\end{eqnarray}
where we have suppressed the superscript $\overline{MS}$ in $\alpha_s$.

\begin{center}\begin{picture}(300,120)(0,0)
\ArrowLine(85,110)(150,90)\ArrowLine(215,70)(150,90)
\ArrowLine(150,90)(215,110)\ArrowLine(150,90)(85,70)
\Vertex(150,90){2}
\Text(150,65)[]{(a)}

\ArrowLine(10,50)(75,30)\ArrowLine(140,10)(75,30)
\ArrowLine(75,30)(140,50)\ArrowLine(75,30)(10,10)
\Photon(42.5,40)(42.5,20){1}{10}
\Vertex(75,30){2}
\Text(75,5)[]{(b)}

\ArrowLine(160,50)(225,30)\ArrowLine(290,10)(225,30)
\ArrowLine(225,30)(290,50)\ArrowLine(225,30)(160,10)
\Photon(257.5,40)(257.5,20){1}{10}
\Vertex(225,30){2}
\Text(225,5)[]{(c)}
\end{picture}

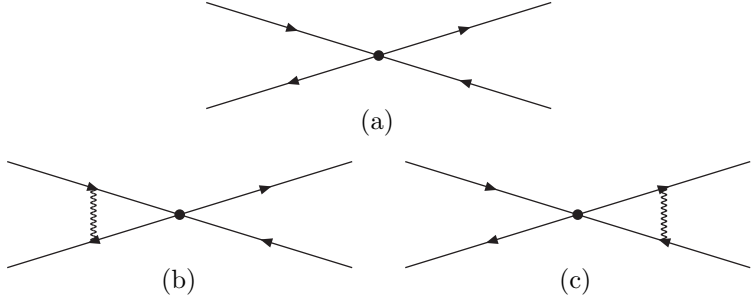
{\sl\bf Fig.6~~Feynman diagrams contributing to $C_{(T^a\otimes T^a) (1\otimes 1)}$
to next-to-leading order of $\alpha_s$ in NRQCD}
\end{center}

In order to determine $Imf_8(^1S_0)$, we must calculate the corresponding
contribution of $\delta{\cal L}_{4-fermion}$ to 
$C^{full~QCD}_{(T^a\otimes T^a) (1\otimes 1)}$
in NRQCD to next-to-leading order in $\alpha_s$.
The relevant Feynman diagrams are shown in Fig.6. They contain a four-fermion
vertex that corresponds to the term $\psi^+T_a\chi\chi^+T_a\psi$ in the
effective Lagrangian. In the limit $v\rightarrow 0$, only Fig.6(b) and Fig.6(c),
which include Coulomb exchange of the gluon, contribute at next-to-leading 
order. Fig.6(a) gives leading order result 
\begin{equation}
Im{\cal M}_{6(a)}=\frac{Imf_8(^1S_0)}{m^2}\xi^{\prime +}
T^a\eta^{\prime}\eta^+T^a\xi.
\end{equation}
The contribution from Fig.6(b) is 
\begin{equation}
Im{\cal M}_{6(b)}=\frac{Imf_8(^1S_0)}{m^2}(C_F-\frac{C_A}{2})
\frac{\pi\alpha_s}{4v}[1-\frac{i}{\pi}(\frac{1}{\epsilon_{IR}}-\gamma_E
+ln4\pi-2ln\frac{mv}{\mu})]
\xi^{\prime +}T^a\eta^{\prime}\eta^+T^a\xi,
\end{equation}
where the imaginary part arises because the incoming quark and antiquark can
scatter on shell before being annihilated by the 4-fermion operator. The 
contribution from Fig.6(c) is
\begin{equation}
Im{\cal M}_{6(b)}=\frac{Imf_8(^1S_0)}{m^2}(C_F-\frac{C_A}{2})
\frac{\pi\alpha_s}{4v}[1+\frac{i}{\pi}(\frac{1}{\epsilon_{IR}}-\gamma_E
+ln4\pi-2ln\frac{mv}{\mu})]
\xi^{\prime +}T^a\eta^{\prime}\eta^+T^a\xi.
\end{equation}
Add the contributions from Fig.6(a),(b),(c) together, we obtain the complete 
result for
$C_{(T^a\otimes T^a)(1\otimes 1)}$ to order of $\alpha_s^3$ in NRQCD
\begin{equation}
\label{NRS}
C^{NRQCD}_{(T^a\otimes T^a)(1\otimes 1)}=\frac{Imf_8(^1S_0)}{m^2}
[1+\frac{\alpha_s}{\pi}(C_F-\frac{C_A}{2})\frac{\pi^2}{2v}].
\end{equation}

Comparing (\ref{PES}) and (\ref{NRS}), we can read off the imaginary 
part of $f_8(^1S_0)$ to next-to-leading order in $\alpha_s$:
\begin{equation}
\label{IMS}
Imf_8(^1S_0)=\frac{(N_c^2-4)\pi\alpha_s^2}{4N_c}[1+\frac{\alpha_s}{\pi}(4b_0ln\frac{\mu}{2m}+A)].
\end{equation}
Note that the factorization approach reproduces the standard prescription
of simply dropping the $1/v$ terms in the perturbatively calculated 
annihilation rate. It is clear that the Coulomb singularity can be factored
into the nonperturbative part trivially in this factorization formula.

Having derived the coefficients $Imf_1(^1P_1)$ and $Imf_8(^1S_0)$, we finally 
come to the overall result for hadronic decay of $1^{+-}$
quarkonium state to next-to-leading order in $\alpha_s$ at leading order
of $v^2$. Substituting (\ref{IMP}) and (\ref{IMS}) into (\ref{WDA}), we get
\begin{eqnarray}\nonumber
\label{all0}
\Gamma(1^{+-}\rightarrow LH)&=&\frac{2(N_c^2-4)C_F\alpha_s^3}
{3N_c^2}(\frac{7\pi^2-118}{48}-ln\frac{\mu}{2m})H_1\\&
+&\frac{(N_c^2-4)\pi\alpha_s^2(\mu)}{2N_c}
(1+\frac{\alpha_s}{\pi}(4b_0ln\frac{\mu}{2m}+A))H_8(\mu).
\end{eqnarray}
Working to all orders in $\alpha_s(\mu)$, the final result is independent of
$\mu$, since the coefficients depend on $\mu$ in such a way that they will cancel
the $\mu$ dependence of matrix elements.

Now we apply our above result to the charmonium system to study the decay
width of $h_c$. In (\ref{all0}) making a choice of $\mu=m_c$ and taking 
$N_c=3$,~~$n_f=3$ we obtain
\begin{equation}
\label{num}
\Gamma(h_c\rightarrow LH)=-0.16\alpha_s^3(m_c)H_1
           +2.62\alpha_s^2(m_c)(1+7.10\frac{\alpha_s(m_c)}{\pi})H_8
\end{equation}
It is interesting to note that the contribution of color singlet component is
negative and the QCD radiative correction from color octet component is very 
large.
The matrix elements $H_1$ and $H_8$ have been defined explicitly in
NRQCD and they are difficult to derive from first principles of QCD. People have tried
to compute them using lattice simulations \cite{lattice}. In practice they can
be determined phenomenologically. Heavy quark spin symmetry provides 
approximate relations between them and the corresponding two parameters in 
the expressions of decay widths of P-wave triplet $\chi_{cJ}(J=0,1,2)$ states. 
At leading
order of $v^2$, they are equal respectively. A rough estimate of $H_1$ and 
$H_8$ have been given in \cite{mang} by comparing the theoretical result of
$\chi_{cJ}$ decay to order $\alpha_s^3$ with experimental data. There they 
don't give the coefficients of $H_8$ at order $\alpha_s^3$, because in
$\chi_{cJ}$ decays, the contributions of color-octet component are the same for 
J=0,1,2, and can be treated as just one parameter. However it is not $H_8$ 
mentioned above. Here, as an approximation, using the estimated value for 
$H_1$ and $H_8$ to the order of $\alpha_s^2$ in \cite{pwave}, 
$$
H_1=15.3\pm 3.7Mev,~~~~H_8=3.26\pm 0.73Mev,
$$
$$
\alpha_s(m_c)=0.25\pm 0.02,
$$
we roughly get $\Gamma=0.80\pm 0.20Mev$. A more reliable estimate will be
obtained with a complete theoretical result for the $\chi_{cJ}$ decay width
to order $\alpha_s^3$.

In this work we use a general factorization formula which is based on NRQCD to
calculate the annihilation rate of $1^{+-}$ quarkonium, we see that the 
infrared divergence appearing in previous calculations can be factored into 
the long-distance perturbative matrix element rigorously. Our result is also 
free from the Coulomb singularity. The corresponding case of $\chi_{cJ}$ 
production
through gluon fragmentation has been studied in \cite{ma}. It is clear from
our calculation that the failure of previous factorization assumption is due
to the fact that only color singlet component was considered and all 
contributions from color octet were neglected. In that sense the previous 
result is incomplete and therefore the
infrared divergence may appear in some cases such as the annihilation and production
of P-wave quarkonium even at leading order in $v^2$. Our calculation shows that 
the rigorous factorization formula can
separate short-distance perturbative effects from long-distance nonperturbative
effects correctly and can therefore provide a systematical calculation for 
quarkonium
decay and production to any order in $\alpha_s$ and in $v^2$, because it is based on a solid theoretical foundation.

\vfill\eject

\vfill\eject

\end{document}